\documentclass[12pt,preprint]{aastex}

\shorttitle{Hidden Trigger for Starburst Arc in M\,83?}
\shortauthors{D\'{\i}az et al.}

\received{2006 June} \accepted{2006 July}
\begin{document}

\title{Hidden Trigger for the Giant Starburst Arc in M\,83?}

\author{Rub\'en J. D\'{\i}az\altaffilmark{1,2}, Horacio Dottori\altaffilmark{3}, Maria P. Aguero\altaffilmark{2}, \\
Evencio Mediavilla\altaffilmark{4}, Irapuan
Rodrigues\altaffilmark{3}, Damian Mast\altaffilmark{2}}

\altaffiltext{1}{Gemini Observatory, Southern Operations Center,
Chile; e-mail:rdiaz@gemini.edu} \altaffiltext{2}{Observatorio
Astron\'omico de C\'ordoba, Universidad Nacional de C\'ordoba and
CONICET,
 Argentina} \altaffiltext{3}{Instituto de F\'{\i}sica, Universidade Federal do
Rio Grande do Sul, Brazil} \altaffiltext{4}{Instituto de Astrofisica
de Canarias, Spain}

\begin{abstract}

The huge star formation events that occur at some galactic centers
do not provide enough clues as to their origin, since the
morphological signatures of the triggering mechanism are smeared out
in the timescale of a few orbital revolutions of the galaxy core.
Our high spatial resolution three-dimensional near-infrared
spectroscopy for the first time reveals that a previously known
hidden mass concentration is located exactly at the youngest end of
a giant star-forming arc. This location, the inferred average
cluster ages, and the dynamical times clearly indicate that the
interloper has left behind a spur of violent star formation in
M\,83, in a transient event lasting less than one orbital
revolution.  The study of the origin (bar funneling or cannibalized
satellite) and fate (black hole merging or giant stellar cluster) of
this system could provide clues to the question of core growing and
morphological evolution in grand-design spiral galaxies. In
particular, our TreeSPH numerical modeling suggests that the two
nuclei could coalesce, forming a single massive core in about 60
million years or less.

This work is based on observations made at the Gemini South
Telescope.

\end{abstract}


%\newpage

\section{INTRODUCTION}

The enormous energy output detected in many cores of galaxies is one
of the key issues in the study of galaxies and their evolution;
notwithstanding, several questions remain unsolved.  Are accretion
onto super-massive black holes and violent star formation just
co-evolving phenomena or necessary partners of the activity?  How is
the detailed physics of the mechanisms triggering the nuclear
extended violent star formation? What is the relationship of the
triggering mechanisms to galaxy evolution? The main challenge facing
these issues is that developed stages of large star formation events
at galactic centers do not provide sufficient clues to their origin,
since the morphological signatures of the triggering mechanism are
smeared out on the timescale of a few orbital revolutions of the
galaxy core.  Here we present the discovery of hidden evolutionary
links in one of the extraordinary transient events in the life of a
galaxy like our own, which occur when the remnant of an accreted
(galactic or extragalactic) body arrives at the galactic nuclear
region. For M\,83, this arrival is accompanied by the fireworks of
the violent star formation arising in the gas-rich environment of
this galaxy, giving us the unique opportunity in the nearby universe
for studying the detailed physics of the so-called nuclear
starbursts and the initial stages of the super-massive black hole
growth in the center of galaxies.

M\,83 is a galaxy with grand-design spiral structure and could be
taken at first glance to be one of the nearest (distance 3.7\,Mpc;
RC3) normal spiral galaxies.  Its central region has progressively
gained attention since it was known to harbor the nearest ``hot
spot'' or Sersic-Pastoriza nucleus (Sersic \& Pastoriza 1965), later
identified as one of the brightest nearby giant HII complexes
(Arsenault \& Roy 1986), and now studied as the nearest nuclear
massive starburst (Fig. 1). It has been subject of several detailed
observational studies, including a wealth of observations with the
Very Large Telescope ({\it VLT}) and the Hubble Space Telescope
({\it HST}), but the dynamical origin of the starburst has remained
elusive. Several hypotheses have been discussed, from the influence
of a not very close companion (Rogstad, Lockart \& Wright 1974) or
the resonance patterns of the global weak bar (Elmegreen, Chromey \&
Warren 1998; Petitpas \& Wilson 1998), down in scale to the presence
of a nuclear bar in the morphology at near- and mid-infrared
spectral ranges (Gallais et al. 1991). Clues to the dynamical
signature of a trigger have been provided initially by long-slit
observations of the nuclear region in a study by Thatte, Tecza \&
Genzel (2000), who report two peaks in the slit profile of the
stellar radial velocity dispersion. The lack of spatially extended
spectroscopic information did not allow them to fix the position and
to confirm the presence of a second nucleus.  This result was
interpreted as revealing the presence of two dynamical centers,
possibly the consequence of the off-center nuclear bar postulated in
previous low-resolution studies (Gallais et al. 1991; Telesco 1988).
However the existence of an off-centered nuclear bar is not expected
from observations and modeling (Maciejewski \& Sparke 2000; Heller
\& Shlosman 1994). The existence of two off-centered nuclei, one of
them coincident with the optical nucleus and the other located a few
arcseconds to the west of the optical nucleus, was proposed (Mast et
al. 2002; Mast, D\'{\i}az \& Ag\"uero 2006) by a team using the
Multi-functional Spectrograph at the Bosque Alegre Astrophysical
Station (D\'{\i}az et al. 1999) to determine the ionized gas radial
velocity and radial velocity dispersion fields from the optical
emission.

A deep study (Harris et al. 2001) of the violent star formation
taking place in M\,83 was made by a thorough photometric analysis of
the 45 most massive clusters in the giant star-forming arc of M\,83
with the Wide Field Planetary Camera 2 of the {\it HST}. This giant
arc is located between 3$\arcsec$ and 7$\arcsec$ from the galaxy
center, spans about 15$\arcsec$ (255\,pc), and includes about 20
massive young clusters similar to 30\,Dor, one of the largest young
clusters in the Local Group of galaxies. By comparing the broad- and
narrow-band photometry with theoretical population synthesis models,
the age and mass of each cluster were estimated (Harris et al.
2001). The main conclusion was that the starburst began about or
less than 10\,Myr ago and that the clusters may dissolve on a
10\,Myr timescale. More recently, Sakamoto et al. (2004) studied the
CO emission in M\,83 with the Submillimeter Array and found that the
distribution and kinematics of the molecular gas is typical for
barred galaxies down to 1\,kpc radii, although they confirm unusual
kinematics around the double nucleus in the central $\sim300$\,pc.
The relatively low spatial resolution of the velocity field
($\sim3\arcsec$) leads these authors to conclude that the second
nucleus would coincide with the center of the bulge. They discuss
the dynamics of the M\,83 central region in the context of the bar
instability and inner Lindblad resonance of the disk and conclude
that the nuclear starburst of M\,83 owes much to the bar-driven gas
dynamics for accumulating molecular gas toward the central 300\,pc.

In order to understand the nature of the double nucleus
configuration and its possible relation to the giant arc of star
formation, minimizing the effects of dust and improving the relative
low spatial resolution ($\sim2\arcsec$) of our previous optical
observations, we applied the new observational techniques of
three-dimensional (3D) spectroscopy at near-infrared (NIR)
wavelengths performed at sub-arcsecond spatial resolution. We
complement these new observations with numerical simulations, which
reveal what can be considered a growing galactic core in a
grand-design spiral galaxy.

\section{OBSERVATIONS}

\paragraph{Observations.} We used the Cambridge Infrared Panoramic Survey
Spectrograph (Parry et al. 2000), built at the Cambridge Institute
of Astronomy, during its visit to the Gemini South 8.1 m telescope
in March 2003. The observations were taken with an Integral Field
Unit (IFU) sampling of 0$\farcs36$ (6.4\,pc) in an elliptical
arrangement with a size of 13$\arcsec$$\times$5$\arcsec$. The array
has 490 hexagonal doublet lenses attached to fibres and provides an
area filling-factor near 100\%.  The IFU was oriented at PA
120$^{\circ}$ (Fig. 1) and was centered in a point midway between
the optical nucleus position and the possible position of the hidden
nucleus previously determined from our optical 2D kinematics (Mast
et al. 2002). The set of 490 spectra covers the spectral range
1.2-1.4\,$\micron$, including the emission lines Pa$\beta$
1.3\,$\micron$ and [FeII] 1.26\,$\micron$, and the spectral
resolution is $\sim$3200. During the observations the peripheral
wave front sensor of Gemini active optics was used, and the achieved
image quality was excellent (FWHM$\approx0\farcs5$); therefore, the
focal plane was somewhat sub-sampled by the used configuration. The
data were reduced using IRAF (distributed by the National Optical
Astronomy Observatory), ADHOC (2D kinematics analysis software
developed by Marseille's Observatory), SAO (spectra processing
software developed by the Special Astrophysical Observatory,
Russia), and standard worksheets and image processing software.  Due
to the complex data output the spectra have been carefully reduced
one by one, using the prominent sky emission lines as wavelength and
profile references. The general techniques used have been previously
described in other works (D\'{\i}az et al. 1999, Mast et al. 2006).
In most of the field the S/N ratio was higher than 10, and the
average radial velocity uncertainty resulted about 6\,km\,s$^{-1}$.
We present here a mean velocity field that includes both nuclei.  We
also constructed the Pa$\beta$ continuum map, which is shown in
Figure\,2 and can be compared for reference with the {\it HST}
pseudo color optical image, combined from F439W, F555W, and F702W
filters, presented in Figure\,1.

\paragraph{Astrometry.}  The position reference system was taken from the two
most recent papers that show optical and NIR images with accurate
astrometry (Thatte et al. 2000; Harris et al. 2001), which have
coordinate system differences of about $0\farcs1$ (determined from
their figures). Using some $J$-band compact features appearing in
Figure\,2 as position references and identifying them in the
previous high-resolution images (mainly the one at F814W band), we
were able to provide coordinates for the galaxy visible nucleus
(defined as the continuum emission peak), the bulge geometrical
center and the dark rotation center reported here. The resulting
coordinates of the visible nucleus are $\alpha=13^h37^m0.95^s$,
$\delta=-29^{\circ}51'55.5''$ (J2000.0) with a 2 $\sigma$
uncertainty of $0\farcs15$. The bulge geometrical center, defined as
the symmetry center of the outer $K$-band isophotes of the galactic
central region (Thatte et al. 2000; central arc-minute, $r<500$\,pc)
is at $\alpha=13^h37^m0.57^s$, $\delta=-29^{\circ}51'56.9''$
(J2000.0), with a 2 $\sigma$ uncertainty of $0\farcs8$ arising
mainly in the photometric determination of this symmetry center.
Finally, the position of the dark rotation pattern center was
determined by fitting a pure rotational model (Satoh disk; Binney \&
Tremaine 1991, pp. 44-45) with varying inclination. The resulting
coordinates are $\alpha=13^h37^m0.46^s$,
$\delta=-29^{\circ}51'53.6''$ (J2000.0), with a 2 $\sigma$
uncertainty of $0\farcs7$.

\section{RESULTS \& DISCUSSION}

We present in Figure\,2 the radial velocity field of the ionized gas
observed at near infrared wavelengths.  An inspection of the
continuum map (with a resolution of $\sim0\farcs6$) and the radial
velocity field clearly shows that the main rotation center is far
($7\farcs8\pm0\farcs7$ or 140$\pm$13\,pc to the WNW,
PA\,$=284^{\circ}\pm5^{\circ}$) from the optical nucleus position,
defined as the maximum peak in the $J$-band continuum emission. The
optical nucleus is also located out of the global symmetry center:
the galactic bulge or central spheroid geometry center is located at
$3\farcs5\pm0\farcs8$ or 63$\pm$14\,pc to the WNW,
PA\,$=249^{\circ}\pm5^{\circ}$, from the visible nucleus position.
The relative positions of the visible nucleus, the bulge geometrical
center, and dark rotation center can be seen in Figures\,2 and 3.
The radial velocity gradient across the optical nucleus implies a
total mass not larger than 10$^7$\,M$_{\odot}$ inside a radius of
2$\arcsec$. In particular, the fit of a Satoh disk model yields a
total mass of $(2\pm1)\times10^6$\,M$_{\odot}$ with an inclination
of $50^{\circ}\pm10^{\circ}$. This dynamical mass value is
consistent with the mass calculated (Thatte et al. 2000) for the
stellar component, using population synthesis models. The main
rotation center has no obvious emitting structure associated in the
continuum map of Figure\,2 and in the  near-IR HST imagery, but is
coincident with the largest lobe in the 10$\micron$ map (Gallais et
al. 1991) shown in Figure\,3, and the second largest lobe in the
6\,cm radio map (Telesco 1988). The rotation pattern has been fitted
with a Satoh disk (Binney \& Tremaine 1991) with inclination
$50^{\circ}\pm15^{\circ}$ and a total mass of
$(16\pm4)\times10^6$\,M$_{\odot}$. Considering the lack of emission
in our 1.3$\micron$ continuum map of this highly obscured region, we
estimate that it should have a mass to light ratio 10 to 100 times
larger than the optical nucleus in the $J$ band. The relatively
large mass of the intruder would explain the optical nucleus
off-centering and perturbed appearance in the high resolution
images. The maximum possible central mass of this dark rotation
center can be estimated by fitting a point-like gravitational source
to the observed radial velocity gradient, inside a $1\arcsec$ radius
from the rotation center and considering a beam smearing of two
sampling elements ($0\farcs72$). Therefore, the largest supermassive
black hole that can be fitted to the unresolved inner region of the
rotation center should have a mass not larger than
$(3\pm1)\times10^6\,($sin$\, i)^{-1}$\,M$_{\odot}$.

The hidden mass concentration is located precisely at the younger
end of the star forming arc, in a region where star clusters with
extinction possibly larger than 10 mag have been reported (Gallais
et al. 1991; Mast et al. 2006). Therefore, the core of the hidden
mass concentration could eventually be explained by the presence of
one or a few super-massive star clusters such as the one forming the
optical nucleus (stellar luminosity corresponding to
$2.5\times10^6$\,M$_{\odot}$; Thatte et al. 2000 ), but highly
obscured in this case.

This obscured rotation center (hereafter intruder nucleus) could
correspond to an interloper gaseous body funneled to the
circumnuclear environment by the global bar dynamics in a scenario
such as that proposed by Elmegreen et al. (1998) for M\,83.  This
kind of scenario has been thoroughly modeled by several teams (e.g.
Heller \& Shlosman 1994; Pinner, Stone \& Teuben 1995), but until
now there was no direct observational evidence of the behavior of
the central region during a bar-fueling event, so the results
presented here could be important for testing the detailed physics
of the models. Currently the best and unbiased (in terms of
extinction) picture of the bar dynamics in M\,83, is the
interferometric imaging of CO emission (Sakamoto et al. 2004). Based
on a relatively low spatial resolution velocity field, these authors
conclude that the second nucleus would coincide with the center of
the bulge and that the optical nucleus could be an interloper. The
dynamics of the M\,83 bar instability and resonances have been
extensively discussed by Sakamoto et al. (2004) and by other authors
previously (e.g. Handa et al. 1990; Kenney \& Lord 1991; Elmegreen
et al. 1998), and the main conclusion that can be drawn is that,
whatever the trigger, the nuclear starburst of M\,83 owes much to
the bar for the accumulation of molecular gas in the central region.
Although the Sakamoto et al. (2004) velocity map has much lower (5-8
times) resolution than the Gemini+CIRPASS data discussed here, a
slight distortion and isovelocities crowding along the minor axis
and northwest from the optical nucleus position could be identified
as the hidden mass concentration.  This concentration is located
just at the edge of a fairly massive molecular gas concentration in
what could be interpreted as bar streaming flow of M\,83, but it
appears strongly asymmetric in intensity and shape with respect to
the global pattern.

It could be possible that the hidden nucleus is part of the material
fed by the bar, but this would imply that the bar is capable of
funneling extended bodies of ten million solar masses and tens of
parsec sizes to the circumnuclear region. Moreover, the bar feeding
scenario is difficult to support due to a couple of details:

1st.- The strong asymmetrical appearance of the mid-infrared (Fig.
3) and radio maps (6\,cm; Telesco 1988), which, in addition to their
tidal shape, would not show a continuity of their features with the
bar dust lanes (marking the main bar streaming regions; Athanassoula
1992).

2nd.- There is at least one other case of a single nucleus off
center with respect to both the geometric and the kinematical
center, in the strongly barred Seyfert galaxy NGC\,1672 (D\'{\i}az
et al. 1999). In both NGC\,1672 and M\,83 the strong offset between
the optical nucleus, the bulge center, and the main kinematic center
would indeed remain as a puzzle in a bar funneling scenario.
Nevertheless, this scenario cannot be ruled out without
high-resolution numerical models.

An alternative would be the arrival of an accreted satellite core
into the highly gaseous circumnuclear environment of M\,83.  As this
is a theoretical scenario less explored at the scale of our
observations, we will thoroughly analyze the possibilities in the
present case of M\,83.

Malin \& Hadley (1997) have found a low-brightness arc outside the
disk of M\,83 and explained it as the remnant of an accreted
satellite.  The global appearance of the HI disk could show some
evidences of a past interaction, such as a large HI tidal feature
that surrounds the optical disk of the galaxy (Park et al. 2001). At
circumnuclear scales, it was found that the integrated H$\alpha$
kinematical profile of the whole complex has a small secondary
Gaussian component, redshifted with respect to the main one; and
infall of material was claimed as a plausible explanation (Arsenault
\& Roy 1986). Moreover, it has been argued (Sofue \& Wakamatsu 1994)
that part of the central region of M\,83 is obscured by a polar dust
lane, which in turn was related to a global warp in the H\,I disk.
Coincidentally or not, the dust lane would end at the galactic
equatorial plane in the region of the hidden mass concentration
discussed here.

There are also a number of galaxies in which the optical nucleus is
off centered with respect to the galaxy center, e.g., NGC\,3227
(Mediavilla \& Arribas 1993), M\,31 (Kormendy \& Richstone 1995),
NGC\,1068 (Arribas, Mediavilla \& Garcia-Lorenzo 1996), and
NGC\,5033 (Mediavilla et al. 2005). In many cases the off centering
has been related to a galaxy merger, and minor mergers have been
theoretically considered in the past as a source of gaseous nuclear
fueling (Taniguchi \& Wada 1996). It is expected in the above
scenario that the nucleus of the cannibalized satellite is accreted
to the central region of the galaxy, where theoretically it should
generate a star formation trail (Saslaw \& De Young 1972; Saslaw
1975) as it perturbs a gas rich environment and eventually produces
a supermassive black hole binary. A statistical analysis of the
previous photometric results (Harris et al. 2001) allows a
quantitative approach to this proposition. We calculated the average
age of the young massive clusters in 25$^{\circ}$ angular sectors of
the 100$^{\circ}$ ringlet (within a radial range from 50 to
150\,pc), and found a clear age gradient across the arc.  This age
gradient is more obvious when the age of the oldest clusters is
considered; in the east extreme of the arc the oldest cluster is
25\,Myr old, while in the northwest extreme, in the region of the
intruder nucleus, the oldest cluster is only 5\,Myr old.  This age
difference is beyond any statistical uncertainty; moreover, the
appearance of the H\,II regions in Figure\,1 shows clearly that the
star-forming regions near the intruder nucleus are more compact and
luminous, and therefore less evolved.  Another striking
characteristic of the giant arc of star formation is that the
location of the youngest clusters and the H$\alpha$ morphology
indicate that the star formation has been radially propagated from
the interior of the arc outward to its radial perimeter.  As has
been stated previously (Harris et al. 2001), it appears that the
5-7\,Myr population has evacuated interstellar material from most of
the active arc and that star formation is continuing along the
radial edges of the region, where the gas density is probably still
high. This phenomenon had no previous explanation, but would
naturally fit in a scenario where star formation has propagated
radially from the path of the intruder nucleus around the galactic
center (see Figs. 1 and 3). This picture is fully consistent with
the dynamical crossing times of the galactic central regions, which
have been estimated using the observed positions of the mass
concentrations, the inferred path of the captured body in the
circumnuclear region and the rotation curve of M\,83 (Ag\"uero \&
D\'{\i}az 2004) from optical, near-infrared, and radio observations.
We consider that the orbit of the intruder mass is partially
described by an ellipse with a semi-major axis not smaller than
120\,pc (as seen in Fig. 1, and with the bulge center in one of the
foci), a range of possible inclinations from the global one
(20$^{\circ}$-30$^{\circ}$) to that inferred at the kinematical
center (40$^{\circ}$-60$^{\circ}$), and the range of possible total
masses enclosed in the radial range from 100 to 200\,pc.  This
Keplerian approach gives a minimum dynamical crossing time of
3\,Myr, and maximum of 12\,Myr, with the most probable value around
5\,Myr (semi-major axis 130\,pc, inclination 45$^{\circ}$, enclosed
mass $2\times10^8$\,M$_{\odot}$), which is fairly consistent with
the age gradient depicted in Figure\,1.

The observational evidences that would therefore support the
proposed capture and star formation triggering picture are:

i) the hidden mass concentration is located at the younger end of
the star-forming arc (Fig. 1);\\ ii) the optical nucleus is out of
the galactic geometry center and seems to have an associated stellar
trail (see Figs. 1 and 3), possibly pointing toward the
galactic center;\\
iii) the continuum maps at radio wavelengths (Rogstad et al. 1974,
Telesco 1988) and at 10$\micron$ (Gallais et al. 1991; see our Fig.
3 for a superposition of the optical and mid-IR features) have a
strong tidal appearance, with the largest emission at mid infrared
wavelengths located precisely at the position of the kinematically
detected dark mass;\\ iv) the mentioned possible off-plane dust lane
related with global H\,I map (Sofue \& Wakamatsu 1994) would end in
the region of the hidden mass concentration;\\ v) the age of the
oldest clusters in the starburst arc is about 25\,Myr (Harris et al.
2001), similar to the dynamical crossing time of the central
kiloparsec ($\sim26$\,Myr);\\ vi) the age difference between the
youngest and the oldest ends of the arc (a few Myr) is similar to
the dynamical crossing time at a hundred parsec scale in M\,83.

\paragraph{Modeling.} In order to make a theoretical approach to
the dynamical evolution of this kind of scenario we simulate the
encounter between the system components following the evolution of
their stellar and gaseous contents. A set of numerical simulations
were done using a TreeSPH code (Hernquist \& Katz 1989). The
circumnuclear galactic potential was represented by a fixed
spherical Hernquist (1990) potential fitted to the most detailed
rotation curve available for M\,83 (Ag\"uero \& D\'iaz 2004), with a
total mass of $1.5 \times 10^{11}$\,M$_\odot$, scale length of
2.5\,kpc, and cutoff radius of 40\,kpc. It was centered at the bulge
geometrical center, defined as the symmetry center of the outer
$K$-band isophotes (see explanation in Section\,2). Nuclei models
(optical nucleus and intruder nucleus) were constructed following
the general prescription for $N$-body galaxy realizations (Hernquist
1993), including gaseous and stellar disks plus a spherical
component. Initial conditions for these models were based on the
observed sizes, masses and 2D radial velocity distributions. A total
of 3000 particles were used for the mass distribution representing
the optical nucleus, arranged in a disk component ($M_{disk} = 0.9
\times 10^6$\,M$_\odot$; scale length $l = 4.2$\,pc; vertical scale
height $h = 0.4$\,pc; Q$_{Toomre} = 1.5$; $V_{circ(peak)} =
25$\,km\,s$^{-1}$) and a spherical component ($M_{sphere} = 1.1
\times 10^6$\,M$_\odot$; core radius $r = 2.1$\,pc; cutoff radius
$r_{cutoff} = 44$\,pc). For the intruder nucleus we used 17,000
particles arranged in a disk component ($M_{disk} = 8.8 \times
10^6$\,M$_\odot$; $l = 6.8$\,pc; $h = 0.68$\,pc; Q$_{Toomre} = 1.5$;
$V_{circ(peak)} = 50$\,km\,s$^{-1}$) plus a spherical component
($M_{sphere} = 7.2 \times 10^6$\,M$_\odot$; $r = 3.4$\,pc; cutoff
radius $r_{cutoff} = 68$\,pc). Considering the observed molecular
gas distribution (Lundgren et al. 2004), a fraction of 10\% of the
optical nucleus mass and 50\% of the intruder dark mass were
assigned to gaseous massive (SPH) particles.

As Figure\,2 shows, Pa$\beta$ spider-like diagrams, characteristic
of a rotating disk, are observed around the kinematical center and
hidden nucleus. Although they only indicate the gas behavior, the
presence of a stellar disk cannot be ruled out, mainly around the
kinematical center. This leads us to consider  models composed of
spheroids and disk in our simulations. Genzel et al. (2001) found
that both nuclei of the major merger Arp 220 retain rotating gas
disks; therefore, the presence of the disk in the intruder nucleus
is physically possible. The real nature of the mass distribution
would be determined when the near-infrared stellar velocity
dispersion maps (currently under elaboration) become available. As a
first approach to the problem, we model the intruder's orbit and the
main galaxy disk coplanar, for the sake of simplicity and
considering the strong star-forming effect on the main galaxy disk.
Indeed, it has been shown in numerical simulations (Walker, Mihos \&
Hernquist 1996) and in the properties of double nuclei in spiral
galaxies (Gimeno, D\'iaz \& Carranza 2004) that most of the accreted
satellites would be moving in the disk of the main galaxy when they
reach the circumnuclear region.

Figure\,4 shows the evolution of the encounter in terms of the
separation between the optical (original?) galaxy nucleus and the
mass barycenter of the captured galaxy; the initial path was
approached with a long parabolic orbit with a perigalactic distance
$q=1$\,kpc. The models show that the main galaxy nucleus and the
bulk of the intruder mass would form a single massive core in
50-70\,Myr, considering the range of uncertainty in the orbit
inclination.  Note that this is, in fact, an upper limit, since in
the present simulation the central potential of M\,83 is modeled by
a rigid potential, and the dynamical friction between it and the
optical and intruder nuclei is not considered. In fully
self-consistent simulations under elaboration, the merging time may
be considerably shorter. The massive core of a few
$10^7$\,M$_{\odot}$ would finally settle as the new nucleus of M\,83
in less than 100\,Myr, implying a net growth of the central galactic
mass. Furthermore, the whole star formation and nuclei merging event
would last less than a global galactic revolution (about 150\,Myr at
the average radius of 5\,kpc). Mergers between spiral galaxies and
their satellites are of broad interest because they can have a
profound influence on the structure of disks in spiral galaxies
through secular processes (Hernquist 1991; Combes 2001) that do not
destroy the galaxy, unlike to those events involving more massive
companions (whose remnants are more akin to elliptical galaxies
rather than any kind of spiral galaxy). Owing to the complexity of
the merger process, quantitative predictions of the effects of
merging are very difficult without appealing to numerical
simulations applied to the study of individual events, which in turn
can be used to constrain the models of spiral galaxy secular
evolution.  The massive core of a few $10^7$\,M$_{\odot}$ that would
finally settle as the new nucleus of M\,83 would represent about
$10^{-3}$\,M$_{acc}$, the mass that would be accreted during a
Hubble time by a spiral galaxy such as our own (Hernquist 1991) or
like M\,83, which would also be dynamically dominated by a massive
halo (Ag\"uero \& D\'{\i}az 2004).

Whatever the origin of the interloper mass, its presence and
star-forming trail are consistent with the suggestion that minor
mergers and bar fueling are the main sources of bulge growth in
spiral galaxies (Combes 2001), which in turn could explain the
correlation between supermassive black hole mass and galactic bulge
mass (e.g. Ferrarese \& Merrit 2000; Tremaine et al. 2002;
M$_{SMBH}\sim10^{-3}$\,M$_{bulge}$). This correlation can be
extrapolated to the Sbc morphological type of M\,83 and is
consistent with the observations presented here: the presence of a
quasar remnant (M$_{SMBH}\sim10^8$\,M$_{\odot}$) can be discarded in
this spiral galaxy from our kinematic data. Notwithstanding, we
might be in the presence of a circumnuclear disk caught in the last
stages of the evolution towards a collapsed object with nonstellar
activity. In this scenario the fate of M\,83 could be to evolve to
an earlier type galaxy, eventually harboring an active nucleus, as
might be expected from the statistical excess of active nuclei in Sa
galaxies (Ho, Filippenko \& Sargent 1997), which also ties bulge
formation and evolution to massive black hole formation and
evolution.

\section{FINAL REMARKS}

Summarizing, several pieces of observational evidence support the
proposed capture and star formation triggering picture: (1) the
hidden mass concentration, detected by 3D near-infrared
spectroscopy, is precisely located at the younger end of the star
forming arc; (2) the optical nucleus is out of the galactic geometry
center and seems to have a stellar trail possibly pointing toward
the geometric center; (3) the continuum maps at radio wavelengths
and at 10$\micron$ have a strong tidal appearance, with the largest
emission at mid-infrared wavelengths located precisely at the
position of the kinematically detected dark mass; (4) a possible
off-plane dust lane related to global H\,I map would end in the
region of the hidden mass concentration; (5) the age of the oldest
clusters in the starburst arc would be about 25\,Myr, similar to the
dynamical crossing time of the central kiloparsec ($\sim26$\,Myr);
(6) the age difference between the youngest and the oldest ends of
the arc (a few Myr) would be similar to the dynamical crossing time
at a hundred parsec scale in M\,83. Therefore, the hidden mass
location and the inferred average cluster ages and the dynamical
times clearly indicate that the interloper has left behind a spur of
violent star formation in M\,83, in a transient event lasting less
than one orbital revolution.

The origin of the interloper could be bar-funneled material or a
cannibalized satellite, and the fate of most of the involved masses
could be a black hole merging or giant stellar cluster.  Whatever
the nature and fate of this complex system, we surely (and luckily)
are witnessing a short event that yields a net growth of the
galactic core in this grand-design spiral galaxy, and the phenomena
reported here should help to make a leap forward in our
understanding of the evolutionary picture of the final stages in a
galactic minor merger or a bar-funneled massive cloud.  We remark
that the reported ages and timescales imply that the observed
phenomenon is extremely short in galactic timescales (probability of
occurrence less than 1\% in a given galaxy), and that its detection
may be unique among the nearby galaxies. The next step in our study
is the execution of higher resolution numerical studies, together
with spectroscopic and imaging observations in the mid-infrared
range, focused on the (dark) largest dynamical center, which so far
appears to be bright at these wavelengths.

\paragraph{Acknowledgements.}
R.D. thanks the support of the Instituto de Astrofisica de Canarias
(Spain) and the hospitality of its researchers.  H.D. acknowledges
support from CNPq (Brazil) and Megalit/Millenium grants.  E.M.
thanks the support of the Euro3D RTN. We wish to thank the
Instrumentation Group at the Institute of Astronomy, Cambridge, for
providing CIRPASS and supporting the observations at the telescope.
The Raymond and Beverly Sackler Foundation and PPARC were
responsible for the funding of CIRPASS.  The Gemini Observatory is
operated by the Association of Universities for Research in
Astronomy, Inc., under a cooperative agreement with the NSF on
behalf of the Gemini partnership: NSF (USA), PPARC (United Kingdom),
NRC (Canada), ARC (Australia), CONICET (Argentina), CNPq (Brazil)
and CONICYT (Chile). The NASA/ESA {\it Hubble Space Telescope} is
operated by AURA under NASA contract NAS 5-26555.  This work also
was partially supported by the CONICET grant PIP 5697.

%\clearpage

%\footnotesize

% FIGURES

\clearpage

\begin{figure}
\includegraphics{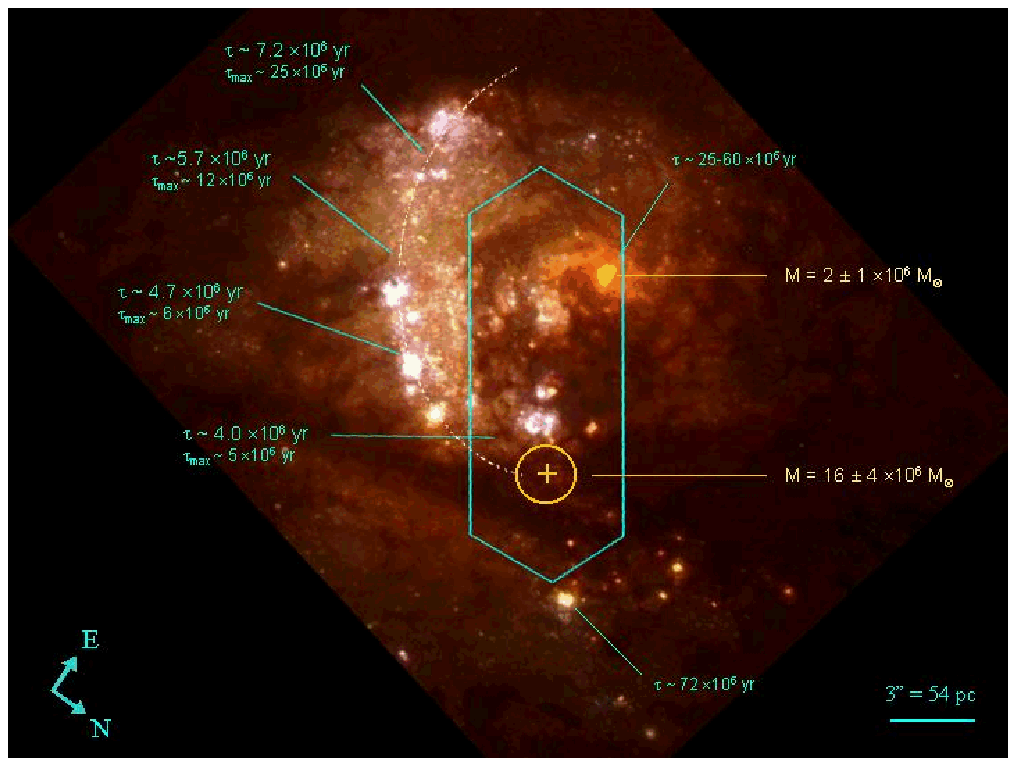} \vspace{15 cm} \caption{{\it HST} false color optical
image, combined from F439W, F555W, and F702W filters.  Point-spread
functions were matched to a common resolution of $0\farcs09$. We
determined the average ages of the young massive star clusters in
the arc (in 25$^{\circ}$ angular sectors) by using the available
data (Harris et al. 2001), which are depicted together with the age
of the oldest clusters in each sector. The integral field observed
with the CIRPASS instrument attached to the Gemini South telescope
has been depicted together with the position of the main rotation
center found (the yellow circle corresponds to the 2 $\sigma$
uncertainty radius).  The red features in the image inside the
integral field area can be compared with the $J$-band continuum
image generated from the spectral data in Fig. 2. Note that the
rotation center (intruder nucleus) is at the youngest end of the
partial ellipse that describes the positions of the main star
forming regions in the giant arc.  Coincidentally, the dynamical
crossing time at this scale is about 5\,Myr. This evidence is
consistent with a trail of violent star formation triggered by the
passage of the intruder nucleus.}
%\label{figure1}
\end{figure}

\clearpage

\begin{figure}
\includegraphics{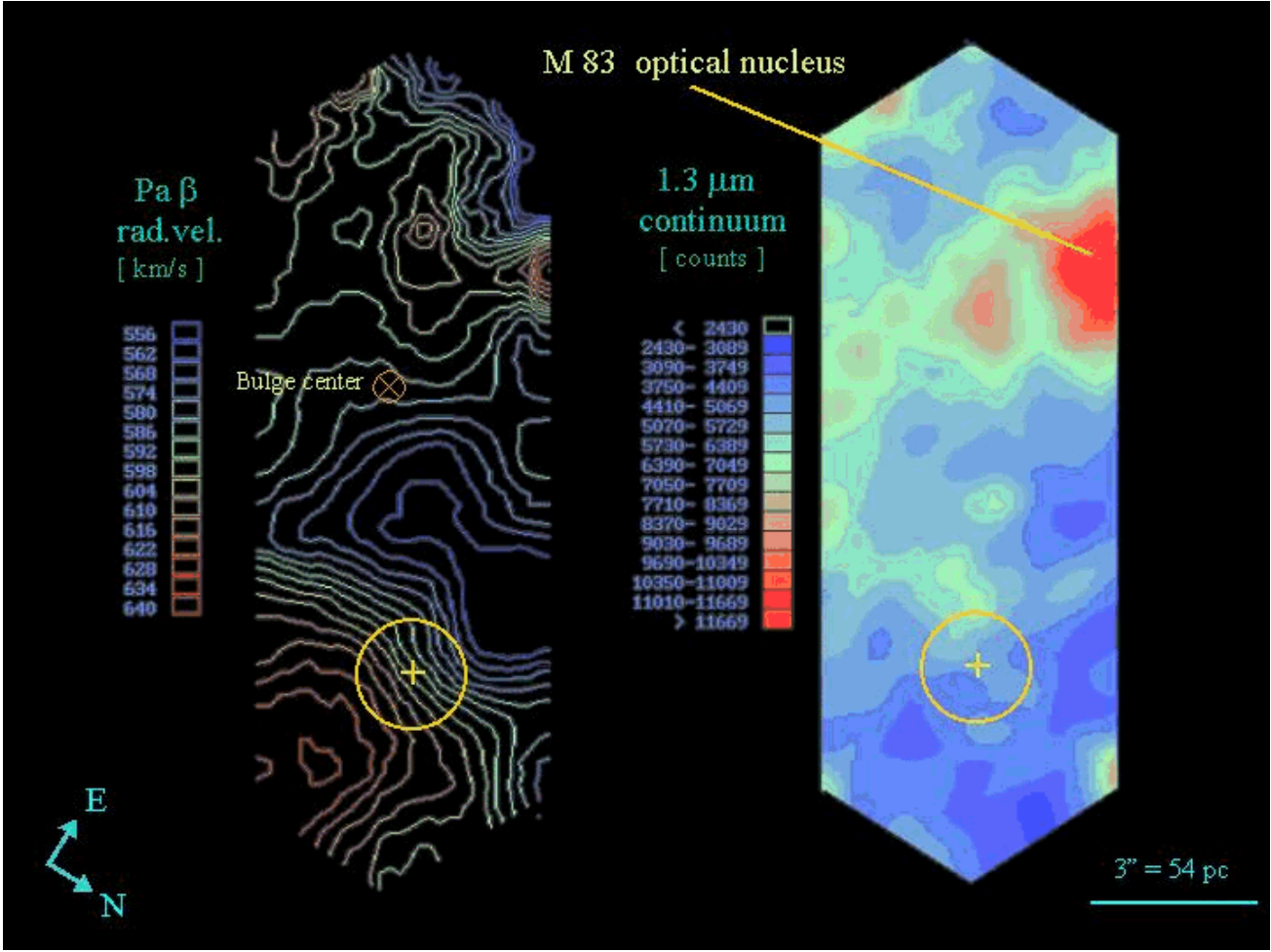} \vspace{15 cm} \caption{{\it Left}: Radial velocity map
of the ionized gas, corresponding to the main integral field
observed with both nuclei.  It corresponds to the field marked in
Fig. 1. The step in isovelocity lines is set equal to the average
uncertainty, but the shape of the field does not qualitatively
change even at a 3 $\sigma$ display.  {\it Right}: Image generated
from the continuum emission in the spectral region 1.28\,$\micron$
(in the photometric $J$-band domain); the achieved resolution is
$0\farcs6$. The main features can be compared for reference with the
reddest features in Fig. 1. The yellow circle corresponds to the 2
$\sigma$ uncertainty radius, and it is evident that the main
rotation center is dark at $J$-band wavelengths and is located far
from the visible nucleus position and from the geometrical center of
the galactic bulge at near-infrared wavelengths, which is marked
with a cross.}
%\label{figure2}
\end{figure}

\clearpage

\begin{figure}
\includegraphics{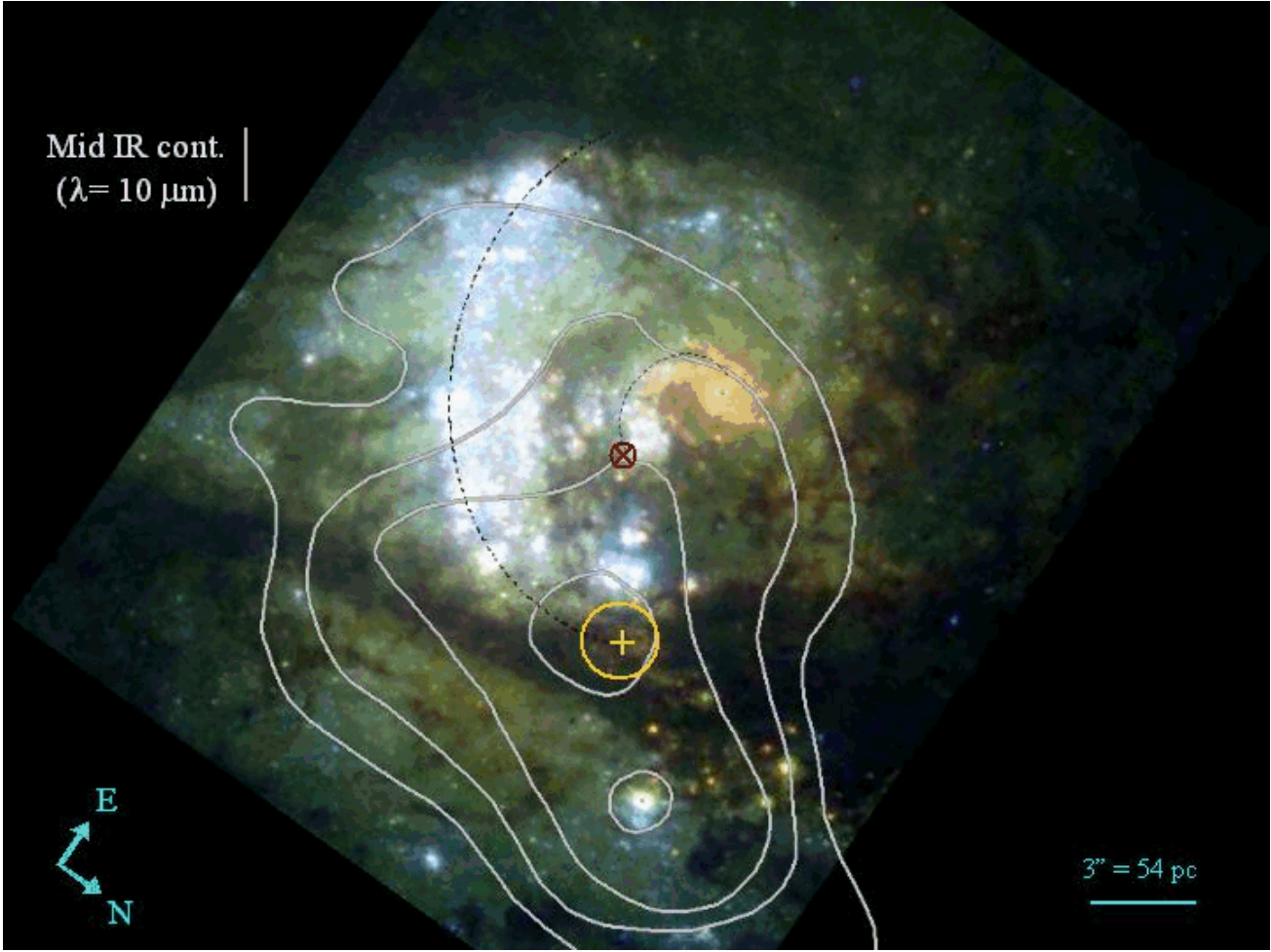} \vspace{15 cm} \caption{Image constructed from the
filters F300W, F547M, and F814W.  In this display the youngest
clusters of the field are enhanced. A strong tidal shape can be seen
in the mid-infrared continuum map at 10\,$\micron$ wavelengths
(Telesco 1988), which is depicted in grey isophotes using the
corresponding position references (Gallais et al. 1991).  The
position uncertainty of the rotation center corresponding to the
intruder nucleus is marked with a circle. Strikingly, this position
is also coincident with the largest lobe in the mid-infrared
emission. Note the partial ellipses that describe the main regions
in the giant arc and a possible stellar trail of the optical
nucleus.}
%\label{figure3}
\end{figure}

\clearpage

\begin{figure}
\includegraphics{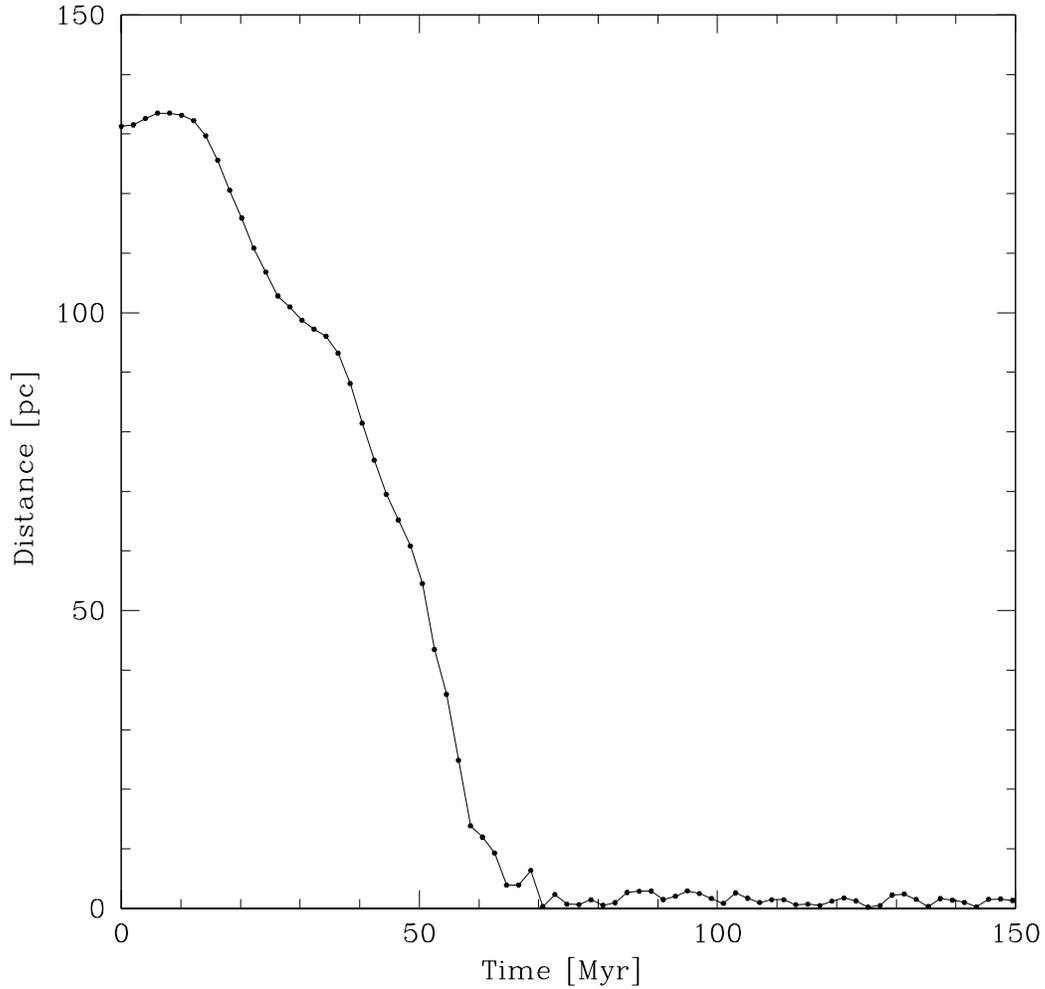}
\vspace{15 cm} \caption{Evolution of the encounter in terms of the
separation between the main galaxy nucleus and the mass barycenter
of the captured body. Time is shown with respect to the present
configuration and the range of uncertainty arises in the unknown
orbit inclination.  It can be seen that the optical original nucleus
of M\,83 and most of the intruder mass would probably form a single
massive core in 50-70\,Myr. This massive core of a few
$10^7$\,M$_{\odot}$ would finally constitute the new nucleus of
M\,83, implying a net growth of the central galactic mass. }
%\label{figure4}
\end{figure}

%\clearpage

\end{document}